\documentclass[sigplan,10pt]{acmart}
\usepackage{pervasives}

\AtBeginDocument{%
  }

\copyrightyear{2024}
\acmYear{2024}
\setcopyright{acmlicensed}\acmConference[PPoPP '24]{The 29th ACM SIGPLAN Annual Symposium on Principles and Practice of Parallel Programming}{March 2--6, 2024}{Edinburgh, United Kingdom}
\acmBooktitle{The 29th ACM SIGPLAN Annual Symposium on Principles and Practice of Parallel Programming (PPoPP '24), March 2--6, 2024, Edinburgh, United Kingdom}
\acmDOI{10.1145/3627535.3638489}
\acmISBN{979-8-4007-0435-2/24/03}

\begin{document}

\title{Fast Kronecker Matrix-Matrix Multiplication on GPUs}

\author{Abhinav Jangda}
\email{ajangda@microsoft.com}
\affiliation{
  \institution{Microsoft Research}
  \city{Redmond}
  \state{Washington}
  \country{USA}
}

\author{Mohit Yadav}
\email{myadav@umass.edu}
\affiliation{
  \institution{University of Massachusetts Amherst}
  \city{Amherst}
  \state{Massachusetts}
  \country{USA}
}

\begin{abstract}
Kronecker Matrix-Matrix Multiplication (Kron-Matmul) is the multiplication of a matrix with the Kronecker Product of several smaller matrices.
Kron-Matmul is a core operation for many scientific and machine learning computations.
State-of-the-art Kron-Matmul implementations utilize existing tensor algebra operations, such as matrix multiplication, transpose, and tensor matrix multiplication.
However, this design choice prevents several Kron-Matmul specific optimizations, thus, leaving significant performance on the table.

To address this issue, we present \sysname{}, an efficient technique for Kron-Matmul on single and multiple GPUs.
\sysname{} is independent of linear algebra operations enabling several new optimizations for Kron-Matmul.
Thus, it performs up to 40.7$\times$ and 7.85$\times$ faster than existing implementations on 1 and 16 GPUs respectively.
\end{abstract}



\maketitle 

\section{Introduction}
\emph{Kronecker Matrix} is a widely used data format in machine learning~\cite{kru, Doping,Kronecker-Graphs, KISS-GP-1, KISS-GP-2, skip, love, mtgpr} and scientific computations~\cite{HyPA, PLOS-Kron, Viljanen2022}.
A Kronecker Matrix is a block matrix of shape $\text{PM}\times \text{QN}$ and is the result of the \emph{kronecker product} of two matrix factors of shape $\text{P}\times \text{Q}$ and $\text{M}\times \text{N}$.
\emph{Kronecker Matrix-Matrix Multiplication} (Kron-Matmul) is the multiplication of an input matrix with a kronecker matrix.
Kron-Matmul is the key operation for computations that represents their data as a Kronecker Matrix.
For example, training Gaussian Processes (GPs), which are a class of machine learning models, involves multiplication of GPs' kernel matrix with the dataset matrix.
In several state-of-the-art GPs~\cite{KISS-GP-1, KISS-GP-2}, the kernel matrix is a kronecker product of smaller matrix factors.
Hence, training of these GPs involves Kron-Matmul of the kernel matrix with the training dataset matrix.

There are two existing algorithms for Kron-Matmul: the shuffle algorithm~\cite{shuffle-algo} and the fused tensor matrix multiply transpose (FTMMT) algorithm~\cite{10.1016/j.cam.2003.10.010}.
Both algorithms run for multiple iterations, where at each iteration, algorithms multiply the input matrix with a factor to generate an intermediate matrix.
This intermediate of an iteration is the input for the next iteration.
These algorithms differ in how they perform the multiplication in each iteration.
The shuffle algorithm uses a series of reshape, matrix multiplication, and transpose operations.
Existing single-GPU implementations of the shuffle algorithm, GPyTorch~\cite{gpytorch} and PyKronecker~\cite{pykronecker}, use NVIDIA cuBLAS~\cite{cublas} for matrix multiplication and generate optimized transpose kernels.
Similarly, the multi-GPU implementation in Cyclops Tensor Framework (CTF)~\cite{ctf} uses distributed matrix multiplication and transpose.
The FTMMT algorithm represents the input matrix as a multi-dimensional tensor and fuses the transpose and multiplication using linear algebra engines, such as COGENT~\cite{tcgen} and cuTensor~\cite{cutensor} for single-GPU and \textsc{Distal}~\cite{distal} for multi-GPU.
Hence, both algorithms use existing linear algebra operations.

However, due to this design choice, these implementations miss several Kron-Matmul specific optimizations leading to the following three inefficiencies.
First, the transpose in the shuffle algorithm is significantly expensive.
Our experiments found that the transpose step can take up to 80\% of the total execution time on both single- and multi-GPU executions.
Second, linear algebra GPU kernels are not optimized for Kron-Matmul.
For example, the matrix multiplication in the shuffle algorithm is performed on small and rectangular matrices, which is an inefficient case for NVIDIA cuBLAS.
Moreover, when caching data from the global to shared memory, COGENT performs a high number of shared memory bank conflicts.
Third, linear algebra operations used in both algorithms require full intermediates in the global memory at each iteration, leading to unnecessary loads and stores from the global memory.
This process can be optimized by fusing multiple linear algebra operations in a single kernel.
However, existing implementations cannot perform this fusion because they use general linear algebra operations instead of Kron-Matmul specific operations.
Similarly, multi-GPU implementations communicate per GPU intermediate at every iteration, leading to high communication volume.

In this paper, we present \sysname{} to address above limitations.
\sysname{}'s Kron-Matmul algorithm is not based on existing linear algebra operations and thus, enables new optimizations for single- and multi-GPU scenarios.
\sysname{}'s algorithm divides rows of the input matrix into slices of size equal to the factor's column and multiplies each slice with all columns of the factor (Section~\ref{sec:new-algo}).
Then the algorithm stores consecutive elements in the intermediate matrix as the multiplication of consecutive slices with same column.
Thus, our algorithm writes output elements at the correct index, removing the need for memory shuffle operations like transpose and reshape.
\sysname{}'s GPU implementation contains a novel tiling methodology that assigns multiple slices and columns to each thread (Section~\ref{sec:new-cuda-impl}).
The implementation caches inputs in the shared memory while minimizing the shared memory bank conflicts and performing coalesced global memory accesses (Section~\ref{sec:gl-load-shmem-store}).
The algorithm also enables us to fuse multiplications with multiple factors in a single GPU kernel by storing intermediates in the shared memory, leading to reduced global memory accesses (Section~\ref{sec:fusion}).
Furthermore, \sysname{}'s multi-GPU implementation minimizes the communication volume by performing multiple local multiplications on each GPU before communicating the intermediate of the last local multiplication (Section~\ref{sec:cuda-multi-gpu}).

\sysname{} provides significant performance speedup over state-of-the-art single and multi-GPU Kron-Matmul implementations. 
On an NVIDIA Tesla V100 GPU, \sysname{} provides up to 40.7$\times$ speedup over GPyTorch~\cite{gpytorch}, 6.40$\times$ over COGENT~\cite{tcgen}, and 5.41$\times$ over cuTensor (Section~\ref{sec:eval:singlegpu}).
On a system with 16 NVIDIA Tesla V100 GPUs, \sysname{} performs 7.85$\times$ better than CTF~\cite{ctf} and 5.33$\times$ better than \textsc{Distal}~\cite{distal} (Section~\ref{sec:eval:multigpu}).
We also integrated \sysname{} into GPyTorch to reduce the training time of several Gaussian Process techniques by up to 6.20$\times$.
\sysname{} is publicly available at \url{https://github.com/abhijangda/fastkron}. 



\begin{figure*}
  \footnotesize
\begin{subfigure}[b]{0.34\textwidth}
$
\mathtt{reshape}(\X_{2\times 4} \rightarrow \X_{4\times 2})\times \F{2}_{2\times 2} = \Y{2}_{4\times 2} =\\
\left[\begin{array}{@{}c@{\hspace{1em}}c@{}}
  x_{11}\times f^2_{11} + x_{12}\times f^2_{21} & x_{11}\times f^2_{12} + x_{12}\times f^2_{22}\\
  x_{13}\times f^2_{11} + x_{14}\times f^2_{21} & x_{13}\times f^2_{12} + x_{14}\times f^2_{22}\\
  x_{21}\times f^2_{11} + x_{22}\times f^2_{21} & x_{21}\times f^2_{12} + x_{22}\times f^2_{22}\\
  x_{23}\times f^2_{11} + x_{24}\times f^2_{21} & x_{23}\times f^2_{12} + x_{24}\times f^2_{22}
\end{array}\right]
$
\caption{Reshape $\X$ and compute $\Y{2} = \X \times \F{2}$\label{fig:kron-matmul:a}}
\end{subfigure}
\begin{subfigure}[b]{0.3\textwidth}
$
\mathtt{trans}(\mathtt{reshape}(\Y{2}_{4\times 2} \rightarrow \Y{2}_{2\times 2\times 2}), 1, 2) = \\
\begin{bmatrix}
  \left[ \begin{array}{@{}c@{\hspace{0.75em}}c@{}}
    \sum_{\ii{} = 1}^{2} x_{1\ii{}} \times f^2_{\ii{}1} & \textcolor{red}{\sum_{\ii{} = 1}^{2} x_{1\ii{}} \times f^2_{\ii{}2}}\\
    \textcolor{red}{\sum_{\ii{} = 3}^{4} x_{1\ii{}} \times f^2_{\frac{\ii{}}{2}1}} & \sum_{i = 3}^{4} x_{1\ii{}} \times f^2_{\frac{\ii{}}{2}2}
  \end{array}\right]\\
  \left[ \begin{array}{@{}c@{\hspace{0.75em}}c@{}}
    \sum_{\ii{} = 1}^{2} x_{2\ii{}} \times f^2_{\ii{}1} & \textcolor{blue}{\sum_{\ii{} = 1}^{2} x_{2\ii{}} \times f^2_{\ii{}2}}\\
    \textcolor{blue}{\sum_{\ii{} = 3}^{4} x_{2\ii{}} \times f^2_{\frac{\ii{}}{2}1}} & \sum_{\ii{} = 3}^{4} x_{2\ii{}} \times f^2_{\frac{\ii{}}{2}2}
  \end{array} \right]
\end{bmatrix}
$
\caption{Transpose last dimensions of $\Y{2}_{2\times 2\times 2}$\label{fig:kron-matmul:b}} 
\end{subfigure}
\begin{subfigure}[b]{0.35\textwidth}
  $
  \mathtt{reshape}(\Y{2}_{2\times 2\times 2} \rightarrow \Y{2}_{2\times 4}) = \Y{2}_{2\times 4} = \\
  \left[\begin{array}{@{}c@{\hspace{0.75em}}c@{\hspace{0.75em}}c@{\hspace{0.5em}}c@{}}
    \sum_{\ii{} = 1}^{2}x_{1\ii{}}\times f^2_{\ii{}1} & \sum_{\ii{} = 3}^{4}x_{1\ii{}}\times f^2_{\frac{\ii{}}{2}1} & \sum_{\ii{} = 1}^{2}x_{1\ii{}}\times f^2_{\ii{}1} & \ldots\\ 
    \sum_{\ii{} = 1}^{2}x_{2\ii{}}\times f^2_{\ii{}1} & \sum_{\ii{} = 3}^{4}x_{2\ii{}}\times f^2_{\frac{\ii{}}{2}1} & \sum_{\ii{} = 1}^{2}x_{2\ii{}}\times f^2_{\ii{}1} & \ldots 
  \end{array}\right]
  $
  \caption{Reshape to $\Y{2}_{2\times 4}$ \label{fig:kron-matmul:c}}
  \end{subfigure}
\caption{\label{fig:kron-matmul}First iteration of the shuffle algorithm for Kron-Matmul of $\X_{2\times 4}$ and $\F{1}_{2\times 2} \kron \F{2}_{2\times 2}$. Reshape transforms shape of a tensor to other shape. Transpose exchanges the elements of two dimensions of a multi-dimensional tensor.}
\end{figure*}
\begin{figure*}
  \footnotesize
  \begin{subfigure}[b]{0.29\textwidth}
  $
  \left[\begin{array}{@{\hspace{0.2em}}c@{\hspace{0.75em}}c@{\hspace{0.75em}}c@{\hspace{0.75em}}c@{\hspace{0.2em}}}
    \tikzmarkin[hor=style green]{x1a}x_{11} & x_{12}\tikzmarkend{x1a} & \tikzmarkin[hor=style green]{x1b}x_{13} & x_{14}\tikzmarkend{x1b}\\
    x_{21} & x_{22} & x_{23} & x_{24}\\
  \end{array} \right]
  \times
  \left[\begin{array}{@{\hspace{0.2em}}c@{\hspace{0.75em}}c@{\hspace{0.2em}}}
    \tikzmarkin[ver=style green]{b1} f^2_{11} & f^2_{12}\\
    f^2_{21}\tikzmarkend{b1} & f^2_{22}\\
  \end{array} \right]
  $
  \caption{Sliced-Multiply 1$^{\text{st}}$ col of $\F{2}$ with 1$^{\text{st}}$ row of $\X$ to get first 2 elements of $\Y{2}$\label{fig:ex-new-algo:a}}
  \end{subfigure}
  \hfill
  \begin{subfigure}[b]{0.29\textwidth}
    $
    \left[\begin{array}{@{\hspace{0.2em}}c@{\hspace{0.75em}}c@{\hspace{0.75em}}c@{\hspace{0.75em}}c@{\hspace{0.2em}}}
      \tikzmarkin[hor=style orange]{x1c}x_{11} & x_{12}\tikzmarkend{x1c} & \tikzmarkin[hor=style orange]{x1d}x_{13} & x_{14}\tikzmarkend{x1d}\\
      x_{21} & x_{22} & x_{23} & x_{24}\\
    \end{array} \right]
    \times
    \left[\begin{array}{@{\hspace{0.2em}}c@{\hspace{0.75em}}c@{\hspace{0.2em}}}
      f^2_{11} & \tikzmarkin[ver=style orange]{b2}f^2_{12}\\
      f^2_{21} & f^2_{22}\tikzmarkend{b2}\\
    \end{array} \right]
    $
    \caption{Sliced-Multiply 2$^{\text{nd}}$ col of $\F{2}$ with 1$^{\text{st}}$ row of $\X$ to get next 2 elements of $\Y{1}$\label{fig:ex-new-algo:b}} 
  \end{subfigure}
  \hfill
  \begin{subfigure}[b]{0.4\textwidth}
    $
    \Y{2} =
    \left[\begin{array}{@{\hspace{0.2em}}c@{\hspace{0.75em}}c@{\hspace{0.75em}}c@{\hspace{0.75em}}c@{\hspace{0.2em}}}
      \tikzmarkin[hor=style green]{r11}\sum_{\ii = 1}^{2}x_{1\ii}\times f^2_{\ii1}\tikzmarkend{r11} & \tikzmarkin[hor=style green]{r12}\sum_{\ii = 3}^{4}x_{1\ii}\times f^2_{\frac{\ii}{2}1} \tikzmarkend{r12}& 
      \tikzmarkin[hor=style orange]{r13}\sum_{\ii = 1}^{2}x_{1\ii}\times f^2_{\ii1}\tikzmarkend{r13} & \tikzmarkin[hor=style orange]{r14}\ldots\tikzmarkend{r14}\\ 
      \tikzmarkin[hor=style green]{r21}\sum_{\ii = 1}^{2}x_{2\ii}\times f^2_{\ii1} \tikzmarkend{r21}& \tikzmarkin[hor=style green]{r22}\sum_{\ii = 3}^{4}x_{2\ii}\times f^2_{\frac{\ii}{2}1} \tikzmarkend{r22} & 
      \tikzmarkin[hor=style orange]{r23}\sum_{\ii = 1}^{2}x_{2\ii}\times f^2_{\ii1} \tikzmarkend{r23}& \tikzmarkin[hor=style orange]{r24}\ldots\tikzmarkend{r24} 
    \end{array}\right]
    $
    \caption{Do steps \textbf{(a)} and \textbf{(b)} with both rows of $\X$ to get $\Y{2}$ 
    \label{fig:ex-new-algo:c}}
  \end{subfigure}
  \caption{First iteration of the \sysname{} Kron-Matmul algorithm of $\X_{2\times 4}$ with $\F{1}_{2\times 2} \kron \F{2}_{2\times 2}$. 
  Elements of $\Y{2}$ with the same color are generated by a column of $\F{2}$ with the same color.
  The result of first iteration, $\Y{2}$, is same as in Figure~\ref{fig:kron-matmul}.
  \label{fig:ex-new-algo}}
\end{figure*}

\section{Kronecker Matrix-Matrix Multiplication}
This section presents existing algorithms for Kronecker Matrix-Matrix Multiplication and their limitations.

A \emph{Kronecker Matrix} of $\textbf{G}_{\FP^{1}\FP^{2}\times \FQ^{1}\FQ^{2}}$ is the result of the \emph{kronecker product} of two matrices $\textbf{F}^{1}_{\FP^{1}\times \FQ^{1}}$ and $\textbf{F}^{2}_{\FP^{2}\times \FQ^{2}}$, such that:
\useshortskip
\begin{equation*}
  \textbf{G}_{\FP^{1}\FP^{2}\times \FQ^{1}\FQ^{2}} = \textbf{F}^{1}_{\FP^{1}\times \FQ^{1}} 
  \kron \textbf{F}^{2}_{\FP^{2}\times \FQ^{2}} = 
\begin{bmatrix}
  f^{1}_{11}\textbf{F}^{2} & \ldots & f^{1}_{1\FQ^{1}} \textbf{F}^{2}\\
  \vdots & \vdots & \vdots \\
  f^{1}_{\FP^{1}1}\textbf{F}^{2} & \ldots & f^{1}_{\FP^{1}\FQ^{1}} \textbf{F}^{2}\\
\end{bmatrix}
\end{equation*}
\useshortskip
We refer to matrices $\textbf{F}^{1}_{\FP^{1}\times \FQ^{1}}$ and $\textbf{F}^{2}_{\FP^{2}\times \FQ^{2}}$ as 
\emph{Kronecker factors} of $\textbf{G}_{\FP^{1}\FP^{2}\times \FQ^{1}\FQ^{2}}$. 
We refer to the operation of multiplying the Kronecker product of $\N$ factors $\F{i}_{\FP^{\text{i}}\times \FQ^{\text{i}}}$ with $\X_{\XM\times \prod_{\text{i}} \FP^{\text{i}}}$ to compute $\Yn_{\XM\times \prod_{\text{i}} \FQ^{\text{i}}}$ as \emph{Kron-Matmul}.
In this section, for simplicity we consider all factors are of the same shape $\FP\times \FQ$. 
A naive algorithm for Kron-Matmul computes the Kronecker matrix and then 
matrix multiply (Matmul) it with $\X$. 
However, this algorithm results in a high complexity of $\complexity{\XM\FP^{\N}\FQ^{\N}}$.
We now present two state-of-the-art algorithms for Kron-Matmul that are faster than the naive algorithm.

\subsection{The Shuffle Algorithm}
The \emph{shuffle algorithm}~\cite{shuffle-algo} avoids computing the Kronecker matrix.
The shuffle algorithm runs for $\N$ iterations from $\N$ to $1$, 
with each iteration performing three steps.
An iteration $i$ generates an intermediate matrix $\Y{i}_{\XM\times \FQ^{\N+1-\ii}\FP^{\ii-1}}$, 
which is also the input for the next iteration.
Figure~\ref{fig:kron-matmul} shows Kron-Matmul of $\X_{\XM\times \FP^2}$ with two factors $\F{1}_{\FP\times \FQ}$ and $\F{2}_{\FP\times \FQ}$.
The first iteration multiplies $\X$ with the last factor $\F{2}$.
First, step \textbf{(a)} reshapes $\X_{\XM\times \FP^2}$ to $\X_{\XM\FP\times \FP}$ and then multiply $\X_{\XM\FP\times \FP}$ with $\F{2}_{\FP\times \FQ}$ to obtain $\Y{2}_{\XM\FP \times \FQ}$ (Figure~\ref{fig:kron-matmul:a}).
Then, step \textbf{(b)} reshapes $\Y{2}_{\XM\FP \times \FQ}$ to $\Y{2}_{\XM \times \FP \times \FQ}$ and transposes the last two dimensions of $\Y{2}_{\XM \times \FP \times \FQ}$ (Figure~\ref{fig:kron-matmul:b}).
Finally, step \textbf{(c)} reshapes $\Y{2}_{\XM \times \FQ \times \FP}$ to $\Y{2}_{\XM \times \FQ\FP}$ (Figure~\ref{fig:kron-matmul:c}).
The next iteration performs above steps with $\Y{2}_{\XM \times \FQ\FP}$ and $\F{1}_{\FP\times \FQ}$ to get the final result of Kron-Matmul, $\Y{1}_{\XM \times \FQ^2}$.
The shuffle algorithm performs $\complexity{\XM\FP\sum_{\ii=1}^{\N}\FQ^{\N-\ii}\FP^{\ii}}$ computations, which is better than the naive approach.

\spara{Limitations}
State-of-the-art GPU based implementations of the shuffle algorithm, GPyTorch~\cite{gpytorch} and PyKronecker~\cite{pykronecker}, uses NVIDIA cuBLAS MatMul in step \textbf{(a)} and an efficient GPU kernel for transposing two inner dimensions of a 3-D tensor for step \textbf{(b)}.
However, this transpose of the 3-D tensor cannot be fused with the Matmul.
Since all transpose steps in the algorithm performs $\complexity{\XM\sum_{\ii=1}^{\N}\FQ^{\N-\ii}\FP^{\ii}}$ memory accesses, both implementations spends up to 80\% of the total time in the transpose.
Moreover, the cuBLAS Matmul is not optimized for multiplying a large skinny and small matrix. 

%
%

\subsection{Fused Tensor-Matrix Multiply Transpose}
We can avoid the expensive transpose operation by representing the input $\X_{\XM\times \FP^{\N{}}}$ and intermediates $\Y{i}_{\XM\times \FQ^{\N+1-\ii}\FP^{\ii-1}}$ as 3 dimensional tensor and fusing the transpose with the computation using tensor multiplication engines~\cite{tcgen, tblis}.
This algorithm, which we call as Fused Tensor-Matrix Multiply Transpose (FTMMT) algorithm~\cite{10.1016/j.cam.2003.10.010} works as follows.
The algorithm goes from \N{} to 1 iterations and in an iteration $i$ multiplies the tensor with the $i^{th}$ factor and transposes the last two dimensions then reshape to the columns of next factor.
Consider Kron-Matmul of $\X_{\XM\times \FP^3}$ with factors $\F{1}_{\FP \times \FQ}$, $\F{2}_{\FP \times \FQ}$, and $\F{3}_{\FP \times \FQ}$.
In the first iteration, we reshape $\X_{\XM\times \FP^3}$ to 3-D tensor $\X_{\XM\times \FP^2 \times \FP}$.
Then, we multiply the last dimension of $\X_{\XM\times \FP^2 \times \FP}$ with the last factor $\F{3}_{\FP \times \FQ}$, to obtain $\Y{3}_{\XM \times \FP^2 \times \FQ}$.
Finally, we transpose the second and last dimensions of $\Y{3}_{\XM \times \FP^2 \times \FQ}$ to $\Y{3}_{\XM \times \FQ \times \FP^2}$ and reshape to $\Y{3}_{\XM \times \FQ\FP \times \FP}$
The second iteration multiplies $\Y{3}_{\XM \times \FQ\FP \times \FP}$ with $\F{2}_{\FP \times \FQ}$ and transposes the first and last dims of the result and reshape to $\Y{2}_{\XM \times \FQ^2 \times \FP}$.
Similarly, the third iteration multiplies with $\F{1}_{\FP \times \FQ}$ to obtain $\Y{1}_{\XM \times \FQ^2 \times \FQ}$ and reshape to $\Y{1}_{\XM \times \FQ^3}$.

\spara{Limitations} Although existing single- and multi-GPU tensor multiplication systems, COGENT~\cite{tcgen}, cuTensor~\cite{cutensor}, and \textsc{Distal}~\cite{distal}, improves over the shuffle algorithm, they do not execute each iteration efficiently and do not optimize across iterations.
For example, COGENT caches data in fast memories using the standard approach, i.e., cache contiguous $\FP$ elements of the last dimension from the shared memory to $\FP$ registers of consecutive threads.
This approach leads to shared memory bank conflicts because every $\FP$ element lies in the same shared memory bank when $\FP$ is a multiple of the number of banks.
Moreover, these systems store the output intermediate of the current iteration in the global memory and load the intermediate in the next iteration, leading to high memory accesses and communication volume.

In summary, existing implementations perform high memory accesses and high communication volume because they do not optimize for Kron-Matmul.


\section{The \sysname{} Algorithm}
\label{sec:new-algo}
This section presents a novel Kron-Matmul algorithm, which enables us to develop new optimizations for Kron-Matmul.

Algorithm~\ref{algo:new-algo} is \sysname{}'s algorithm for Kron-Matmul.
For brevity, we present the algorithm for the common case in our evaluation dataset, where all Kronecker factors are of the same shape.
However, it is straightforward to generalize this algorithm to factors of different shape.
The algorithm works as follows.
First, the algorithm allocates two intermediate matrices and set the number of cols of input matrix for the first iteration (line~\ref{line:new-algo:interm-y}--\ref{line:new-algo:k}).
These intermediates are swapped after every iteration.
The algorithm starts the multiplication from the last factor (lines~\ref{line:new-algo:for-f-begin}--\ref{line:new-algo:for-f-end}).
For each factor, the algorithm first computes number of cols of the output intermediate (line~\ref{line:new-algo:l}).
Then the algorithm performs \emph{Sliced Multiply} for each row of $\X$ with all columns of $\F{i}$, where it divides the row into slices of size $\FP$ (line~\ref{line:new-algo:rowSlice}) and multiplies each slice with each column of $\F{i}$ (line~\ref{line:new-algo:accum}).
Then, the algorithm writes the result of each slice and column in $\Y{1}$ (line~\ref{line:new-algo:write-m}).
Finally, the algorithm returns the final result (line~\ref{line:new-algo:ret}). 
The algorithm performs $\complexity{\XM\FP\sum_{\ii=1}^{\N}\FQ^{\N-\ii}\FP^{\ii}}$ computations and $\complexity{\XM\sum_{\ii=1}^{\N}\FQ^{\N-\ii}\FP^{\ii}}$ memory accesses.
Hence, the ratio of computations to memory accesses is $\FP$.

Figure~\ref{fig:ex-new-algo} shows an example of Kron-Matmul of $\X_{2\times 4}$ and $\F{1}_{2\times 2} \kron \F{2}_{2\times 2}$ using \sysname{}'s algorithm.
First, both columns of $\F{2}$ are multiplied with each slice of each row of $\X$ (Figure~\ref{fig:ex-new-algo:a}).
Two elements generated by sliced multiplication of the first column of $\F{2}$ are stored as the first two elements of the first row of $\Y{2}$.
Then, elements generated by sliced multiplication of the second column of $\F{2}$ are stored as the third and fourth elements of the first row of $\Y{2}$ (Figure~\ref{fig:ex-new-algo:b}).
Finally, the intermediate result with the current factor, $\Y{2}$, is used as the input matrix for the next factor.

\begin{algorithm}[t]
  \small
  \caption{The \sysname{} Kron-Matmul algorithm\label{algo:new-algo}}
\begin{algorithmic}[1]
  \State{\textbf{Input}: Matrix $\X_{\XM \times \FP^{\N}}$ and $\N$ Kronecker Factors $\F{i}_{\FP \times \FQ}$}
  \State{\textbf{Output}: Result of Kron-Matmul of $\X$ with all $\N$ $\F{i}$s}
  \State{$\textbf{Y}^1$ and $\textbf{Y}^2$ are new matrices of size $\XM\times \max_{\text{f}=0}^{\N}(\FQ^{\N-\text{f}}\FP^{\text{f}})$}\label{line:new-algo:interm-y}
  \State{$\Y{1}$ = X} \Comment{Copy X to $\Y{1}$}
  \State{K = $\FP^{\N}$} \label{line:new-algo:k}\Comment{Input Intermediate No. of Cols}
  \For{f = $\N \to 1$} \label{line:new-algo:for-f-begin}
   \For{i = $1 \to \XM$} \label{line:new-algo:for-n-begin}
    \State{L = (K$\div \FP$) $\times \FQ$} \label{line:new-algo:l}\Comment{Output Intermediate No. of Cols}
      \For{j = $1 \to \text{L}$} \Comment{Sliced-Multiply $\X$ row and $\F{f}$ col} \label{line:new-algo:for-m-begin}
        \State{rowSlice = (j $\times$ P)$\mod$K} \label{line:new-algo:rowSlice}
        \State{kCol = (j $\div$ P$^{\N-1}$)$\mod$ P} \label{line:new-algo:kcol}
        \State{acc = 0} \label{line:new-algo:acc-0}
        \For{k = $1 \to \FP$}\Comment{Sliced Multiply Accumulate}
          \State{acc += $\Y{0}$[i][rowSlice + k] $\times$ $\F{f}$[k][kCol]} \label{line:new-algo:accum}
        \EndFor
        \State{$\Y{1}$[i][j] = acc} \label{line:new-algo:write-m}
      \EndFor \label{line:new-algo:for-m-end}
   \EndFor \label{line:new-algo:for-n-end}
   \State{$\Y{1}, \Y{2}$ = $\Y{2}, \Y{1}$} \Comment{Swap intermediates}
   \State{K = L}
  \EndFor \label{line:new-algo:for-f-end}
  \State{\Return{$\Y{1}$}} \label{line:new-algo:ret}
\end{algorithmic}
\end{algorithm}

\spara{Comparison with Existing Algorithms}
There is a key difference that separates \sysname{}'s algorithm and MatMul.
In MatMul, consecutive elements in a row of the output are the result of multiplication of consecutive columns of the second matrix with the same row of the first matrix.
However, in \sysname{}'s algorithm, consecutive elements are obtained by multiplying consecutive slices of rows with the same column of the factor.
Thus, \sysname{}'s algorithm writes output elements at the right index, removing any need for extra memory operations, like transpose.



\begin{figure}[t]
\begin{lstlisting}[language=C,
mathescape=true, numbers=left, numberstyle=\footnotesize,
columns=fixed,
    basicstyle=\footnotesize\ttfamily,
    keywordstyle=\textcolor{blue},
    commentstyle=\color{commentgreen},
    firstnumber=1,
  escapeinside={(*}{*)}, % if you want to add comments in code
  morekeywords={blockDim, bid, tid, warpSize, shared,
                syncthreads},
          escapechar=|, xleftmargin=1.0ex,numbersep=4pt]
Ks = (Slices = ($\tile{K}$/P))*$\tile{P}$;|\label{cuda-impl:rowslices}|
shared T Xs[$\tile{M}$][Ks], Fs[$\tile{P}$][$\tile{Q}$]; |\label{cuda-impl:sh-mems:begin}| |\label{cuda-impl:sh-mems:end}|
register T Yr[$\tile{M}$][$\regtile{K}$][$\regtile{Q}$]={0};|\label{cuda-impl:yreg}|
//Compute Element Locations in Y
yQ = (tid.x / Slices) * $\regtile{Q}$; |\label{cuda-impl:yrgrp}|
yK = (tid.x % Slices) * $\regtile{K}$; |\label{cuda-impl:yrblk}|
for($\mathtt{t_P}$ = 0; $\mathtt{t_P}$ < P; $\mathtt{t_P}$ += $\tile{P}$){|\label{cuda-impl:loop-tilep-begin}|
 /*Step 1: Global to Shared Memory*/
 ShiftGToS(Xg, Xs, K, $\tile{M}$, $\tile{P}$, Ks, $\regtile{K}$);|\label{cuda-impl:xsh}| 
 DirectGToS(Fg, Fs, Q, $\tile{P}$, $\tile{Q}$);|\label{cuda-impl:fsh}|
 syncthreads();|\label{cuda-impl:sync-1}|
 for($\mathtt{r_P}$ = 0; $\mathtt{r_P}$ < $\tile{P}$; $\mathtt{r_P}$ += $\regtile{P}$){|\label{cuda-impl:loop-regtilep-begin}|
  register T Xr[$\tile{M}$][$\regtile{K}$][$\regtile{P}$], Fr[$\regtile{P}$][$\regtile{Q}$];
  /*Step 2: Shared to Registers*/
  ShiftSToR(Xs, Xr, $\mathtt{r_P}$, yK, $\regtile{K}$, $\regtile{P}$, $\tile{M}$, $\tile{P}$);|\label{cuda-impl:xreg}|
  DirectSToR(Fs, Fr, $\mathtt{r_P}$, $\tile{P}$, $\tile{Q}$, $\regtile{Q}$);|\label{cuda-impl:freg}|
  /*Step 3: Multiply Accumulate*/
  for(m = 0; m < $\tile{M}$; m++) for(k = 0; k < $\regtile{K}$; k++)
  for(q = 0; q < $\regtile{Q}$; q++) for(p = 0; p < $\regtile{P}$; p++)
    Yr[m][k][q] += Xr[m][k][p] * Fr[p][q];|\label{cuda-impl:accum}|
} syncthreads();|\label{cuda-impl:sync-2}|} |\label{cuda-impl:loop-tilep-end}|
/*Step 4: Registers to Global Memory*/
for(r = 0; r < $\tile{M}$; r++) |\label{cuda-impl:storey-begin}|
 for(b = 0; b < $\regtile{Q}$; b++) for(e = 0; e < $\regtile{K}$; e++){
  yRow = r + bid.x*$\tile{M}$; |\label{cuda-impl:yrow}|
  yCol = (yQ + b) * ($\tile{K}$/P) + yK + e;|\label{cuda-impl:ycol1}|
  yCol = (yCol / ($\tile{K}$/P)) * (K/P) + 
         bid.y * ($\tile{K}$/P) + yCol % ($\tile{K}$/P); |\label{cuda-impl:ycol2}|
  Y[yRow * K + yCol] = Yr[r][b][e];} |\label{cuda-impl:storey-end}|
\end{lstlisting}
\caption{\sysname{}'s \textsc{SlicedMultiplyKernel} for $\X_{\XM\times \XK}$ and $\Fn_{\FP\times \FQ}$ to compute $\Yn_{\XM\times \frac{\XK\FQ}{\FP}}$.
\texttt{Shift*} and \texttt{Direct*} transfers data from global/shared memory to shared memory/registers.
\label{fig:cuda-impl}}
\end{figure}

\begin{figure*}
  \begin{subfigure}[b]{0.48\textwidth}
  \includegraphics{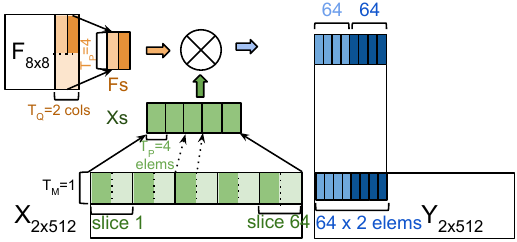}
  \caption{Thread block 0 is assigned to 1$^{\text{st}}$ row of $\X{}$ and 2 cols of $\Fn$ to produce $\frac{512}{8} \times 2$ = 128 elements of $\Y{}$.
  The thread block load 4 elements of all 64 slices into \texttt{Xs} and of columns into \texttt{Fs}, and 
  multiplies each slice and column to produce partial 128 elements of $\Y{}$.
  Then, the thread block moves to the next 4 elements, and updates partial elements to get its final elements.
  ~\label{fig:cuda-impl:tb-tile}}
  \end{subfigure}
  \hspace{1em}
  \begin{subfigure}[b]{0.48\textwidth}
    \includegraphics{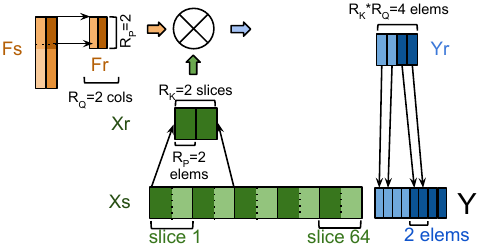}
    \caption{Thread 0 is assigned to first 2 slices of \texttt{Xs} and all 2 cols of \texttt{Fs} to produce 4 elements of \texttt{Yr}. 
    A thread loads 2 elements of both slices in \texttt{Xr} and of cols in \texttt{Fr}, and multiplies each slice and column to produce 4 partial elements.
    Then, the thread moves to next 2 elements and updates partial elements to get final elements.
    Elements for col 1 are stored at index 0 and for col 2 are stored at index 64 of $\Y{}$. 
    ~\label{fig:cuda-impl:th-tile}}
  \end{subfigure}
  \caption{\sysname{}'s tiling to sliced-multiply $\X_{2\times 512}$ and $\Fn_{8\times 8}$ to produce $\Yn_{2\times 512}$ with $\tile{M} = 1, \tile{K} = 512, \tile{Q} = 2, \tile{P} = 4, \regtile{P} = 2, \regtile{Q} = 2,\regtile{K} = 2$. There are $\frac{512}{8} = 64$ slices for each $\X$ row. The CUDA kernel is invoked with $\left\{\frac{2}{1}, \frac{512}{512}, \frac{8}{2}\right\}$ threadblocks.
  \texttt{Xs} and \texttt{Fs} are shared memory buffers. \texttt{Xr}, \texttt{Fr}, and \texttt{Yr} are register buffers.
   ~\label{fig:ex-cuda-impl}}
\end{figure*}
\section{\sysname{}'s CUDA Implementation}
\label{sec:new-cuda-impl}
This section presents \sysname{}'s CUDA kernel with an efficient shared memory caching technique (Section~\ref{sec:gl-load-shmem-store}) and fusion of multiple sliced multiplications (Section~\ref{sec:fusion}).

\sysname{} provides Python and C++ APIs for Kron-Matmul for several data types, including  \texttt{float} and \texttt{double}.
All the API functions call into a type generic C++ implementation of Algorithm~\ref{algo:new-algo}.
The implementation executes the loop of lines~\ref{line:new-algo:for-f-begin}--\ref{line:new-algo:for-f-end} of Algorithm~\ref{algo:new-algo} and returns the output.
The implementation invokes \textsc{SlicedMultiplyKernel} (in Figure~\ref{fig:cuda-impl}) to sliced-multiply $\X_{\XM\times \XK}$ and $\Fn_{\FP\times \FQ}$ to generate $\Yn_{\XM\times \frac{\XK\FQ}{\FP}}$.
The kernel takes global memory addresses for $\X$, $\Fn$, and $\Y{}$, along with their shapes.
Each thread block of the kernel performs the following steps:
\begin{enumerate}[leftmargin=*,labelsep=2pt]
  \item Load slices of rows of $\X{}$ and cols of $\Fn$ into shared memory.
  \item Load part of slices from the shared memory to registers.
  \item Perform sliced multiply accumulate on register buffers to compute multiple elements of $\Y{}$.
  \item When elements of $\Y{}$ are computed, transfer elements from registers to global memory.
\end{enumerate}
The above workflow is similar to NVIDIA CUTLASS~\cite{cutlass} and BLIS~\cite{blis} but with differences in the shared memory caching and element to thread assignment optimized for Kron-Matmul.
We now explain above steps in Figure~\ref{fig:cuda-impl} using an example workflow for $\Yn_{2\times 512}$, $\X_{2\times 512}$, and $\Fn_{8\times 8}$ in Figure~\ref{fig:ex-cuda-impl}.

\spara{Thread Block Tiles}
Each thread block sliced multiply $\left\{\tile{M}, \tile{K}\right\}$ block of $\X_{\XM\times \XK}$ with $\tile{Q}$ cols of $\Fn_{\FP \times \FQ}$ to produce a block of $\left\{\tile{M}, \left(\frac{\tile{K}}{\FP}\times \tile{\FQ}\right)\right\}$ of $\Y{}$.
Thus, the kernel is invoked with $\left\{\frac{\texttt{\XM}}{\tile{\XM}}, \frac{\texttt{\XK}}{\tile
{\XK}}, \frac{\texttt{\FQ}}{\tile{\FQ}}\right\}$ threadblocks.
Each thread produces $\regtile{K} \times \regtile{Q}$ elements for each $\tile{M}$ row.
Thus, each thread block contains $\frac{\tile{K} \div \FP}{\regtile{K}}\times \frac{\tile{\FQ}}{\regtile{Q}}$ threads.
In Figure~\ref{fig:cuda-impl:tb-tile}, each threadblock sliced-multiply $\tile{P} \times \tile{K} = 512$ block of $\X_{2\times 512}$ with $\tile{Q} = 2$ cols of $\Fn_{8\times 8}$ to produce $\frac{512}{8}\times 2 = 128$ elements of a row of $\Yn_{2\times 512}$.

\spara{Global to Shared Memory}
Each thread block caches $\tile{P}$ elements of slices of rows of $\X{}$ and cols of $\Fn$ in the shared memory (lines~\ref{cuda-impl:xsh}--\ref{cuda-impl:fsh} in Figure~\ref{fig:cuda-impl}).
The main loop of the kernel iterates over all $\tile{P}$ tiles and multiplies $\tile{P}$ elements of slices and cols to produce elements of $\Y{}$ (lines~\ref{cuda-impl:loop-tilep-begin}--\ref{cuda-impl:loop-tilep-end}).
In Figure~\ref{fig:cuda-impl:tb-tile}, the thread block caches $\tile{P} = 4$ elements of each slice and col in the shared memory.
Section~\ref{sec:gl-load-shmem-store} presents our efficient shared memory access approach to minimize bank conflicts.

\spara{Shared Memory to Registers}
A thread loads $\regtile{K}$ slices of $\X{}$ and $\regtile{Q}$ cols of $\Fn$ from the shared memory to registers (lines~\ref{cuda-impl:xreg}--\ref{cuda-impl:freg} in Figure~\ref{fig:cuda-impl}).
These slices and columns are multiplied to compute $\regtile{K} \times \regtile{Q}$ elements of $\Y{}$.
Therefore, higher values of $\regtile{K}$ increases the reuse of cols of $\Fn$ and higher values of $\regtile{Q}$ imply higher reuse of slices of $\X$.
Figure~\ref{fig:cuda-impl:th-tile} shows that each thread loads $\regtile{P}$ = 2 elements of $\regtile{K} = 2$ slices and of $\regtile{Q} = 2$ columns to register buffers (\texttt{Xr} and \texttt{Fr}).

\spara{Elements to Thread Mapping}
Each thread computes $\regtile{K} \times \regtile{Q}$ elements of $\Y{}$ stored in registers.
The thread controls the computation intensity using $\regtile{P}$, i.e., each thread loads and multiplies $\regtile{P}$ elements of $\regtile{K}$ slices of $\X$ and $\regtile{Q}$ cols of $\Fn$ (lines~\ref{cuda-impl:loop-regtilep-begin}--\ref{cuda-impl:loop-tilep-end} in Figure~\ref{fig:cuda-impl}).
In Figure~\ref{fig:cuda-impl:th-tile}, with $\regtile{K} = \regtile{Q} = 2$, two slices are multiplied with two cols to obtain 4 elements of \texttt{Y}.

\spara{Registers to Global Memory}
After computing its elements, each thread stores these elements to $\Y{}$ (line~\ref{cuda-impl:storey-begin}--\ref{cuda-impl:storey-end} in Figure~\ref{fig:cuda-impl}).
Since consecutive elements of $\Y{}$ are obtained by multiplying slices of $\X{}$ with the same col of $\Fn$, all $\regtile{K}$ elements are stored consecutively.
Moreover, a group of $\regtile{K}$ elements for a col $c$ are stored at the address starting from $c \times \frac{\XK}{\FP}$.
Therefore, the thread computes its index in $\Y{}$ (lines~\ref{cuda-impl:yrgrp}--\ref{cuda-impl:yrblk} in Figure~\ref{fig:cuda-impl}) and write elements.
In Figure~\ref{fig:cuda-impl:th-tile}, the $\regtile{K} = 2$ elements for the first col of $\Fn$ are stored at indices starting from 0, while the elements for the second col are stored at indices $1 \times \frac{512}{8} = 64$.

\subsection{Efficient Data Movement}
\label{sec:gl-load-shmem-store}
This section describes \sysname{}'s \emph{shift caching} that minimizes shared memory bank conflicts in our algorithm.

The standard approach, which we call \emph{direct caching}, transfers data from the global to shared memory by assigning consecutive threads to contiguous elements and loads contiguous elements from the shared memory to registers of consecutive threads.
This method is used in Matmul and tensor contractions of CUTLASS~\cite{cutlass} and COGENT~\cite{tcgen}.
However, using this method, in our case, leads to high shared memory bank conflicts.
In Figure~\ref{fig:cuda-impl:th-tile}, as $\tile{P}=4$, thread 0 load elements 0--3 of slice 0 stored in \texttt{Xs[0]} to \texttt{Xs[3]}, and thread 1 load elements 0--3 of slice 2 stored in \texttt{Xs[8]} to \texttt{Xs[11]}.
If the number of banks, i.e., \warpSize{}, is 4, then all elements at \texttt{Xs[0]}, \texttt{Xs[8]}, \ldots, \texttt{Xs[64]} fall into the same bank leading to 4 conflicts for every read.

\sysname{}'s \emph{shift caching} performs coalesced global memory accesses and minimizes shared memory bank conflicts to $\lceil\frac{\texttt{WarpSize}}{\texttt{$\tile{P}$}}\rceil$.
Figure~\ref{fig:cuda-impl:gltoshmem} shows the implementation of the shift caching.
When loading from the global to shared memory \texttt{ShiftGToS} finds the slice index for each element and shifts the element forward by the ratio of the slice index and the number of slices per thread ($\regtile{K}$) (line~\ref{gltosh:elem}).
Consequently, when loading elements from the shared memory to registers, \texttt{ShiftSToR} takes the starting slice index of the thread, divides it by the number of slices per thread to obtain the shift, and shifts the element back in the thread's register tile (lines~\ref{shtoreg:shift}--\ref{shtoreg:store-xr}).
In Figure~\ref{fig:cuda-impl:th-tile}, thread 1 stores slice 2 to the shared memory by shifting elements forward by $\frac{2}{\regtile{K}} = 1$, i.e., elements 0--2 are stored at \texttt{Xs[9]}--\texttt{X[11]}, and element 3 is stored at \texttt{Xs[8]}.
Similarly, when storing slice 4, thread 2 shift all elements forward by $\frac{4}{\regtile{K}} = 2$ and stores element 0 of this slice at \texttt{Xs[14]}.
When loading to registers, thread 1 loads element 0 of slice 2 from \texttt{Xs[9]} while thread 2 loads element 0 of slice 4 from \texttt{Xs[14]}.
If \warpSize{} is 4, \texttt{Xs[9]} and \texttt{Xs[14]} fall in different banks, avoiding any bank conflicts.
Section~\ref{sec:eval:micro} shows the effectiveness of the shift method in reducing bank conflicts over the direct method in COGENT~\cite{tcgen} and cuTensor~\cite{cutensor}.

\begin{figure}[t]
  \begin{lstlisting}[language=C,
    mathescape=true, numbers=left, numberstyle=\footnotesize,
    columns=fixed,
        basicstyle=\footnotesize\ttfamily,
        keywordstyle=\textcolor{blue},
        commentstyle=\color{commentgreen},
        firstnumber=1,
      escapeinside={(*}{*)}, % if you want to add comments in code
      morekeywords={blockDim, bid, tid, warpSize, shared,
                    syncthreads},
              escapechar=|, xleftmargin=1.0ex,numbersep=4pt]
ShiftGToS(Xg, Xs, K, $\tile{N}$, $\tile{P}$, Ks, $\regtile{K}$){
 for(m = 0; m < $\tile{M}$; m++) |\label{gltosh:iter-row}|
 for(k = tid; k < Ks; k += bdim) {|\label{gltosh:iter-m}|
  elem = k%$\tile{P}$;|\label{gltosh:elem}| slice = k/$\tile{P}$; |\label{gltosh:slice}|shift = slice/$\regtile{K}$; |\label{gltosh:shift}|
  col = slice*$\tile{P}$ + (elem + shift)%$\tile{P}$; |\label{gltosh:sh-col}|
  Xs[m][col] = Xg[(m + bid.x) * K + k];}}|\label{gltosh:store-xs}|

ShiftSToR(Xs, Xr, $\mathtt{r_P}$, yK, $\regtile{K}$, $\regtile{P}$, $\tile{M}$, $\tile{P}$){ 
 for(m = 0; m < $\tile{M}$; m++) for(q = 0; q < $\regtile{Q}$; q++){|\label{shtoreg:iter-row}| |\label{shtoreg:iter-b}|
  slice = (yK + q);|\label{shtoreg:slice}| shift = (yK / $\regtile{K}$)%$\tile{P}$;|\label{shtoreg:shift}|
  for (p = 0; p < $\regtile{P}$; p++) {|\label{shtoreg:iter-p}|
   elem = $\mathtt{r_P}$ + p;|\label{shtoreg:elem}| round = (elem + shift)%$\tile{P}$;|\label{shtoreg:round}|
   Xr[m][q][p] = Xs[m][slice*$\tile{P}$ + round];}}}|\label{shtoreg:store-xr}|
\end{lstlisting}
\caption{\sysname{}'s shift caching method.
\texttt{ShiftGToS} caches from global to shared memory.
\texttt{ShiftSToR} caches from shared memory to registers.\label{fig:cuda-impl:gltoshmem}}
\end{figure}

\subsection{Fusing Consecutive Sliced Multiplications}
\label{sec:fusion}

This section describes \sysname{}'s fusion mechanism that perform multiple sliced multiplications in a single kernel leading to significant decrease in global memory accesses.

The linear algebra kernels used by existing Kron-Matmul implementations require inputs in the global memory.
Thus, these implementations store the output intermediate of each multiplication in the global memory and load the intermediate again for the next multiplication.
Since \sysname{} is independent of linear algebra kernels, it can fuse consecutive sliced multiplications and store intermediates in the shared memory, thereby, avoiding expensive global memory accesses.
We now discuss the working of fused kernel.


%
%
%

\begin{figure}
  \includegraphics[scale=0.9]{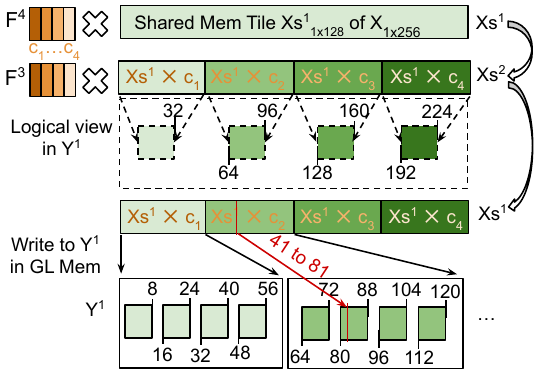}
  \caption{Workflow of the fused kernel for Kron-Matmul of $\X_{1\times 256}$ with 4 factors $\Fn_{4\times 4}$ and $\tile{K} = 128$ by the first thread block.
  The kernel fuses $\Fused = 2$ sliced multiplications (max is 3).
A thread block sliced multiply all 4 cols of $\F{4}$ and $\F{3}$ with $\tile{K}$ shared memory elements and store them in another shared memory buffer.
After each multiplication, the corresponding indices in global intermediate ($\Y{1}$) are shown.
Finally, the output of 2nd sliced multiply is written to $\Y{1}$.\label{fig:fusion}}
\end{figure}

The fused kernel sliced multiplies all cols of $\Fused$ factors with $\tile{K}$ elements and stores the intermediate in shared memory.
Hence, the algorithm using the fused kernel runs for $\lceil \frac{\N}{\Fused}\rceil$ iterations.
After every sliced multiply the number of shared memory elements that are contiguous in the full global memory intermediate reduces by the factor of $\FP$.
Figure~\ref{fig:fusion} shows that the fused kernel for Kron-Matmul of $\X_{1 \times 256}$ and $\Fn_{4 \times 4}$ with $\tile{K} = 128$, sliced multiplies $\Fused = 2$ factors to generate $\tile{K}$ elements in the shared memory for each factor.
After the first sliced multiply, there are $\tile{\FQ}^{1}=4$ sets of $\frac{\tile{K}}{\FP^{1}} = 32$ elements of shared memory with a stride of 32 in the global intermediate.  
After the second multiply, we get $\tile{\FQ}^{2}=16$ sets of $\frac{\tile{K}}{\FP^{2}}= 8$ contiguous elements with a stride of 8 in the global intermediate.
In general, after the $i^{th}$ sliced multiply, there are $\tile{Q}^{\ii}$ sets of $\frac{\tile{K}}{\FP^{\ii}}$ contiguous elements with a stride of $\frac{\tile{K}}{\FP^{\ii}}$ in the global intermediate.
Thus, the fused kernel can compute a maximum of $\Fused = \lfloor\log_\FP \tile{K}\rfloor$ consecutive sliced multiplications.
Moreover, the fusion is valid when all elements of all slices can be stored in the shared memory, i.e., $\tile{P} = \FP$.
Our experiments found this is true for $\FP \leq 32$ and $\FQ \leq 32$.

After the last multiply, the fused kernel transfers elements from the shared to global memory using \texttt{StoreFusedShMem} function (Figure~\ref{fig:cuda-impl:fusedShmemToGL}).
Consider storing element 41 of the shared memory to global memory in Figure~\ref{fig:fusion}.
The function first computes (i) the number of slices in global and shared memory, i.e., 64 and 32 in our example, 
and (ii) the slices of fusion of $\Fused$ factors in the global and shared memory, i.e., 16 and 8 in our example (lines~\ref{line:sh-to-gl:slices}--\ref{line:sh-to-gl:fuse-slices}).
The function now iterates on all $\tile{K}$ elements of each row in the shared memory and store them to global memory using the below steps:
1) Compute the slice of the element in the shared memory tile and scale it to the global memory (line~\ref{line:sh-to-gl:gl-slice}), i.e., 64 in our example;
2) Compute the fused slice index in the shared memory tile and scale to the global memory (line~\ref{line:sh-to-gl:gl-fusedslice}), i.e., 16 in our example;
3) Compute the element index within the fused slice (line~\ref{line:sh-to-gl:gl-elem}), i.e., 1 in our example;
4) Finally, store the element at the sum of the above indices (line~\ref{line:sh-to-gl:store}), i.e., at 81.
Section~\ref{sec:eval:micro} shows that fusion is a key optimization for small $\FP$.

\begin{figure}[t]
  \begin{lstlisting}[language=C,
    mathescape=true, numbers=left, numberstyle=\footnotesize,
    columns=fixed,
        basicstyle=\footnotesize\ttfamily,
        keywordstyle=\textcolor{blue},
        commentstyle=\color{commentgreen},
        firstnumber=1,
      escapeinside={(*}{*)}, % if you want to add comments in code
      morekeywords={bdim, bid, tid, warpSize, shared,
                    syncthreads},
              escapechar=|, xleftmargin=1.0ex,numbersep=4pt]
StoreFusedShMem(Xg, Xs, K, P, Q, $\tile{m}$, $\tile{K}$, $\Fused$) {
 XgSlics = K/P; XsSlics = $\tile{K}$/P;|\label{line:sh-to-gl:slices}|
 XgFuseSlics=K/P**$\Fused$;XsFuseSlics=$\tile{K}$/P**$\Fused$;|\label{line:sh-to-gl:fuse-slices}|
 for(e = tid; e < $\tile{m}$ * $\tile{K}$; e += bdim) { 
  m = e/$\tile{K}$; c = e%$\tile{K}$ |\label{shtoreg:iter-row}|
  //Scale Shared Mem Slice Idx to Global Mem Idx       
  slice = (c/XsSlics)*XgSlics; |\label{line:sh-to-gl:gl-slice}|
  //Scale Shared Fused Slice to Global Mem
  fusedSlice = ((c%XsSlics)/XsFuseSlics) *|\label{line:sh-to-gl:gl-fusedslice}|
                XgFuseSlics;
  //Elem Idx in Fused Slice
  elem = bid.y * XsFuseSlics + c % XsFuseSlics;|\label{line:sh-to-gl:gl-elem}|
  //Column index in Global Memory
  col = slice + fusedSlice + elem;
  Yg[(m+bid.x*$\tile{M}$) * K/P*Q + col] = Xs[m*$\tile{M}$ + c];|\label{line:sh-to-gl:store}|}
  \end{lstlisting}
  \caption{Write output of $\Fused$ fused sliced multiply kernels from shared to global memory.
  \texttt{X**Y} represents $\mathtt{X^Y}.$\label{fig:cuda-impl:fusedShmemToGL}}
  \end{figure}

\subsection{Autotuning Kernel Parameters}
\label{sec:autotuning}
This section describes \sysname{}'s autotuning mechanism to find efficient tile sizes for any Kron-Matmul shape.

Kron-Matmul can be performed on matrices of diverse shapes.
However, there is no single set of tile size parameter values that performs efficiently for all shapes.
Therefore, \sysname{} performs auto-tuning over a range of tile size parameter values for the given shape.
The auto-tuning phase considers all combinations of following values of tile size parameters till the maximum shared memory usage and registers per thread is reached:\\
\textbf{Thread Block Tile Sizes}
(i) $\tile{\XK} \in$ all multiples of $\FP$ till $\XK$,
(ii) $\tile{P} \in$ all factors of $\FP$,
(iii) $\tile{Q} \in$ all factors of $\FQ$, and
(iv) even values of $\tile{M}$ until the number of threadblocks executing in parallel by all SMs reaches a
maximum value.\\
\textbf{Thread Tile Sizes} 
(i) $\regtile{P} \in$ all factors of $\tile{P}$,
(ii) $\regtile{Q} \in$ all factors of $\tile{Q}$,
and (iv) $\regtile{K} \in$ all factors of $\frac{\tile{K}}{\tile{P}}$.\\
This narrowing down of tile size choices based on the available resource usage and the 
occupancy significantly decreases the search space of all choices.
\sysname{} compile CUDA kernels for all combinations of the above tile sizes in parallel and find the kernel with the least execution time.



\section{Distributed Kron-Matmul}
\label{sec:cuda-multi-gpu}

\begin{figure*}
  \includegraphics[scale=0.75]{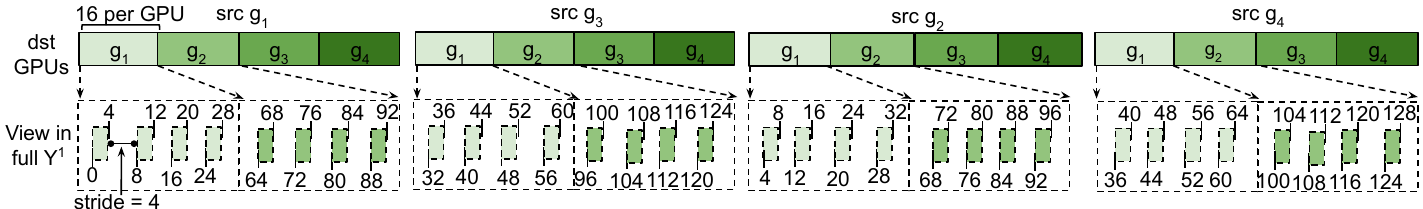}
  \caption{The element distribution of local intermediates of all 4 GPUs for Kron-Matmul on $\X{}_{1\times 256}$ with 4 factors $\Fn_{4\times 4}$ with $\left\{\GPUM, \GPUK\right\} = \left\{1, 4\right\}$.
  Each GPU computes $\Local = 2$ (max value is 3) sliced multiplications locally on the block size $\frac{\XM}{\GPUM} \times \frac{\XK}{\GPUK} = 1\times 64$.
  After $\Local = 2$ multiplications, each local intermediate stores $\frac{\XK \div \GPUK}{\GPUK} = 16$ elements for all 4 GPUs and $\frac{\XK \div \GPUK}{\FP^{\Local}} = 4$ contiguous elements with stride of 4 in the full distributed intermediate.
  \label{fig:distributed}}
\end{figure*}


Existing distributed implementations of the shuffle algorithm in CTF~\cite{ctf} and the FTMMT algorithm using \textsc{Distal}~\cite{distal}, executes each iteration by (i) dividing the input among all GPUs,
(ii) performing multiplications on each GPU to generate its local intermediate, and (iii) communicate intermediates among all GPUs to obtain a globally distributed intermediate as the output of iteration.
This section presents \sysname{}'s distributed Kron-Matmul that minimizes the communication by (i) performing multiple $\Local$ sliced multiplications on each GPU to generate intermediates local to each GPU and (ii) communicating local intermediates to obtain the globally distributed output intermediate of $\Local$ iterations.

\sysname{} performs distributed Kron-Matmul on a homogenous 2D grid of $\left\{\GPUM, \GPUK\right\}$ GPUs by dividing the computation into a block of size $\frac{\XM}{\GPUM} \times \frac{\XK}{\GPUK}$ per GPU.
Since each factor $\Fn_{\FP\times \FQ}$ is much smaller than $\X_{\XM\times \XK}$, \sysname{} requires that all factors are accessible on all GPUs.
Algorithm~\ref{algo:multigpu} is \sysname{}'s distributed Kron-Matmul algorithm, which is executed by each GPU.
The algorithm assumes that all factors are of the same shape, but it is straightforward to support the general case.
First, the algorithm computes GPU block size, allocates intermediate matrices on each GPU, and computes $\Local$ (lines~\ref{line:multi-gpu-algo:tile}--\ref{line:multi-gpu-algo:B}).
Then each GPU can perform $\Local = \log_\FP \frac{\XK}{\GPUK}$ local sliced multiplications before communicating local intermediates to obtain the globally distributed intermediate (line~\ref{line:multi-gpu-algo:sliced-multiply}).
Now the layout of each column of local intermediate is such that (i) a column stores $\GPUK$ parts of size $\frac{\XK \div \GPUK}{\GPUK}$, where the $i^{th}$ local part is stored on the $i^{th}$ GPU of the globally distributed intermediate and 
(ii) a column contains $\frac{\XK \div \GPUK}{\FP^{\Local}}$ elements that lie apart by the same value in the global intermediate.
Figure~\ref{fig:distributed} shows an example of this layout on 4 GPUs.
The algorithm relocates elements on each GPU by sharing these parts among all GPUs (line~\ref{line:multi-gpu-algo:send} and~\ref{line:multi-gpu-algo:recv}) and storing the received elements to correct place using \textsc{StoreGPUTile} function (line~\ref{line:multi-gpu-algo:store}), which is similar to \textsc{StoreFusedShMem}.
Therefore, the algorithm communicates exactly $\GPUM \times \left(\frac{\N \times \GTile{M} \times (\XK{} - \GTile{K})}{\log_\FP \GTile{K}}\right)$ number of values.
\sysname{} uses NVIDIA NCCL~\cite{nccl} for \textsc{Send} and \textsc{Recv}.
If all NVIDIA GPUs in the same $\gM$ supports Point-to-Point accesses, \sysname{} implement lines~\ref{line:multi-gpu-algo:communicate-start}--\ref{line:multi-gpu-algo:send} in a single CUDA kernel, which is more efficient than NCCL.

Similar to SUMMA~\cite{summa}, we divide the process grid into $\{\sqrt{G}, \sqrt{G}\}$, where G is the number of GPUs.
If G is not a perfect square, we set the grid to $\{2^{\lceil\log_2 \sqrt{G}\rceil}, 2^{\lfloor\log_2 \sqrt{G}\rfloor}\}$.
Although complex partitioning approaches for distributed Matmul exists~\cite{cosma, edgar25D, recursive}, our experiments found that our partitioning approach performs well.

\begin{algorithm}[t]
  \small
\caption{Multi-GPU Kron-Matmul using $\GPUM\times \GPUK$ GPUs\label{algo:multigpu}}
\begin{algorithmic}[1]
  \State{Current GPU $\left\{\gM, \gK\right\}$ with in the grid of $\left\{\GPUM, \GPUK\right\}$ GPUs}
  \State{$\GTile{M}, \GTile{K} = \frac{\XM}{\GPUM}, \frac{\text{K}}{\GPUK}$}\Comment{Tile computed by each GPU}\label{line:multi-gpu-algo:tile}
  \State{$\Y{1}$ and $\Y{2}$ are new matrices of size $\GTile{M}\times \GTile{K}$}
  \State{$\Local{}$ = $\lfloor \log_\FP{} \GTile{K} \rfloor$}\label{line:multi-gpu-algo:B}
  \State{$\Y{1} = \X[\gM\times\GTile{M}:(\gM+1)\times\GTile{M}][\gK\times\GTile{K}:(\gK+1)\times\GTile{K}]$}
  \For {f = $\N \to 1$ with step $\Local$} 
    \For {b = $0 \to \Local{}-1$} \Comment{Do $\Local{}$ multiplies per GPU}
    \label{line:multi-gpu-algo:sliced-multiply}
      \State{$\Y{2} = $ \textsc{SlicedMultiplyKernel}$(\Y{1}_{\GTile{M}\times \GTile{K}}, \F{f - b}_{\FP \times \FQ})$}
      \State{$\Y{1}, \Y{2} = \Y{2}, \Y{1}$}
    \EndFor
    \For {dst = $1 \to \GPUK$} \Comment{Share result among GPUs with the same $\gM$}
    \label{line:multi-gpu-algo:communicate-start}
      \If {dst = $\gK$}
        \For{src = $1 \to \GPUK$}
          \State{$\Y{1}_{\text{src}}$ = $\Y{1}[1:\GTile{M}]$[src$\times$$\frac{\GTile{K}}{\GPUK}$:(src+1)$\times$$\frac{\GTile{K}}{\GPUK}$]}
          \State{$\Y{2}_{\text{dst}}$ = $\Y{2}[1:\GTile{M}]$[dst$\times$$\frac{\GTile{K}}{\GPUK}$:(dst+1)$\times$$\frac{\GTile{K}}{\GPUK}$]}
          \If{src $\neq$ dst}
            \State{$\Y{1}_{\text{src}}$ = \textsc{Recv}($\gM$, src)} \label{line:multi-gpu-algo:recv}
          \EndIf
          \State{\textsc{StoreGPUTile}($\Y{2}_{\text{dst}}$, $\Y{1}_{\text{src}}$, $\XK$, $\FP$, $\FQ$, $\GTile{M}$, $\GTile{K}$, $\Local{}$)} \label{line:multi-gpu-algo:store}
          \
        \EndFor
      \Else
        \State{$\Y{1}_{\text{dst}}$ = $\Y{1}[1:\GTile{M}]$[dst$\times$$\frac{\GTile{K}}{\GPUK}$:(dst+1)$\times$$\frac{\GTile{K}}{\GPUK}$]}
        \State{\textsc{Send}($\Y{1}_{\text{dst}}$, $\gM$, dst)} \label{line:multi-gpu-algo:send}
      \EndIf
    \EndFor
    \State{$\Y{1}, \Y{2} = \Y{2}, \Y{1}$}
  \EndFor
  \State{\Return{$\Y{1}$}}
\end{algorithmic}
\end{algorithm}

\section{Evaluation}
\label{sec:eval}
In this section, we evaluate \sysname{} against state-of-the-art implementations of the shuffle algorithm and the FTMMT algorithm on diverse Kron-Matmul sizes.

\spara{Experimental Setup} We run our experiments on a single NVIDIA DGX-2 machine, which contains dual 24-core Intel Xeon CPUs and 16 NVIDIA Tesla V100 GPUs connected using NVLINK 2.
Each Tesla V100 GPU contains 32 GB of global memory, and provides 15.7 Tera Floating Point Operations per Second (TFLOPS) for float and 7.8 TFLOPS for double. 
We use CUDA 12.2 on Ubuntu 22.04 and report the average TFLOPS of 100 runs after a warmup of 10 runs.

\begin{figure}[t]
  \includegraphics[scale=0.7]{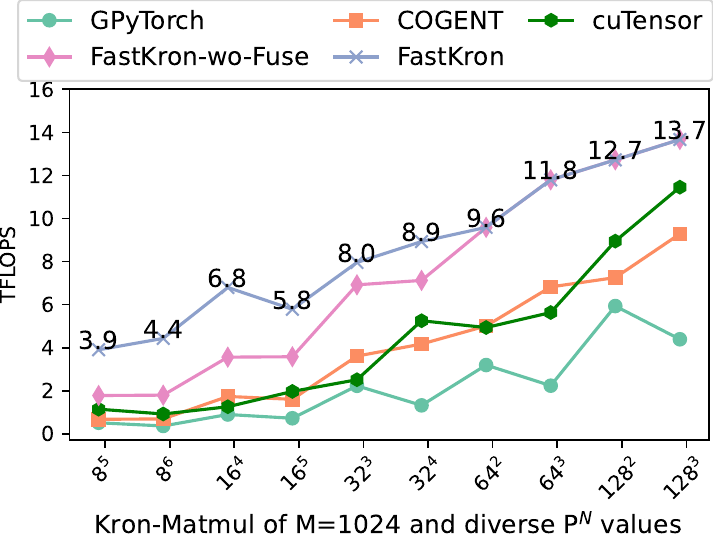}
  \caption{Performance of GPyTorch, COGENT, \sysname{}, and \sysname{} without fusion for float with $\XM = 1024$, $\FP$ = 8 to 128, and the two largest allocatable values of $\FP^\N$.
  \label{fig:results:single-gpu-flops}}
\end{figure}

\subsection{Autotuning Time}
We perform a search over the tile size parameters for each Kron-Matmul problem size to obtain the best performing kernel parameters (Section~\ref{sec:autotuning}).
The autotuner generates upto 10,000 configurations for each problem size.
By compiling kernels in parallel, the auto-tuner takes less than 2 minutes to find the fastest kernel.

\subsection{Single GPU Evaluation}
We first evaluate \sysname{} on a single GPU using microbenchmarks and then on a real world dataset.

\subsubsection{Evaluation Systems}
We perform experiments on the following systems:

\spara{GPyTorch}~\cite{gpytorch} and PyKronecker~\cite{pykronecker} are two state-of-the-art GPU based implementations of the shuffle algorithm. 
Since both implementations call into NVIDIA cuBLAS for Matmul and generate identical CUDA kernels for the transpose, both implementations perform within 10\% of each other. 
Thus, we use GPyTorch 1.11 as the baseline for the shuffle algorithm.

\spara{COGENT}~\cite{tcgen} is a state-of-the-art GPU code generator for tensor contractions.
It fuses the transpose with the multiplication and 
generates optimized code and tile sizes for the FTMMT algorithm for the given Kron-Matmul shape.

\spara{cuTensor}~\cite{cutensor} is a state-of-the-art library for tensor contractions by NVIDIA.
It fuses transpose with multiplications and autotunes at runtime over several tile sizes for the FTMMT algorithm.
We use cuTensor as baseline because we found it performs as good as  manually tuned CUTLASS~\cite{cutlass}.


\spara{\sysname{}} with all optimizations.

\spara{\sysname{}-wo-Fuse} is \sysname{} without the fusion of consecutive sliced multiplications (Section~\ref{sec:fusion}).

\subsubsection{Microbenchmarks}
\label{sec:eval:micro}
We now present results for Kron-Matmul of $\X_{\XM \times \FP^\N}$ with $\N$ factors $\Fn_{\FP \times \FP}$.
Figure~\ref{fig:results:single-gpu-flops} shows the performance of each system with
\XM=1024, the power of two values of \FP, and several values of $\N$. 
The performance improves with the increase in $\FP{}$ and $\N{}$ because (i) the ratio of computation to memory accesses, $\FP$ increases, and (ii) large $\N$ increases the amount of parallelism.
For the largest size, \sysname{} achieves 87\% of the maximum FLOPS of the GPU.

\spara{Impact of Fusion} The fusion optimization improves the performance by 2.20$\times$ for 8$^5$ to 1.15$\times$ for 32$^3$.
Since the shared memory is limited, the number of fused sliced multiplications decreases with an increase in $\FP$, leading to less improvement with increasing $\FP$.
The auto-tuner finds $\tile{P}= 32$ for $\FP{} \geq 64$, hence, fusion is not applied to $\FP{} \geq 64$.

\spara{Speedup over GPyTorch}
\begin{table}[t]
\small
\begin{tabular}{|c|c|r|r|r|r|r|r|}
  \hline
  $\FP$ & $\N$ & \multicolumn{3}{c|}{GPyTorch (ms)} & COGENT &\sysname{}\\
  \cline{3-5}
  & & Matmul & Trans. & Total & \multicolumn{1}{c|}{(ms)}& \multicolumn{1}{c|}{(ms)}\\
  \hline
  8  & 6 & 26 & 45  & 71.0 & 36.4 & 5.76\\
  16 & 5 & 64 & 169 & 238 & 104 & 29.7\\
  32 & 4 & 44 & 159 & 203 & 64.4 & 38.8\\
  64 & 3 & 8.7& 36  & 45.7  & 14.8 & 8.74\\
  \hline
\end{tabular}
\caption{Execution time of GPyTorch's Matmul and transpose, COGENT, and \sysname{} for float with $\XM$ = 1024 and largest values of $\FP$ and $\N$ on a 32GB GPU.
\label{tab:results:times}}
\end{table}
\sysname{} provides a speedup of 7.62$\times$ for $8^5$ to 3.11$\times$ for $128^3$ over GPyTorch because:
(i) \sysname{} avoids the transpose step of GPyTorch.
Table~\ref{tab:results:times} shows that the transpose step takes up to 80\% of the total time of GPyTorch; and
(ii) \sysname{} performs better than cuBLAS Matmul for small and rectangular shapes. 
Table~\ref{tab:results:times} shows that \sysname{} is 0.95$\times$--4.51$\times$ faster than cuBLAS.

\spara{Speedup over COGENT and cuTensor}
Both COGENT and cuTensor provides similar performance.
\sysname{} provides a speedup of 6.40$\times$ for $8^5$ to 1.47$\times$ for $128^3$ over COGENT.
Similarly, \sysname{} provides a speedup of 3.32$\times$ for $8^5$ to 1.2$\times$ for $128^3$ with maximum speedup of 5.41$\times$ at $16^4$.
This speedup is because:
(i) \sysname{} fuse consecutive sliced multiplications and store intermediates in the shared memory;
and (ii) the shift caching method decreases shared memory bank conflicts over COGENT and cuTensor, which uses the direct caching method.
For example, Table~\ref{fig:sh-txs} shows that \sysname{} generates up to 3.10$\times$ less load and 3.18$\times$ less store transactions than COGENT.

\begin{table}[t]
  \small
\begin{tabular}{|c|c|r|r|r|r|r|r|}
  \hline
  $\FP$ & $\N$ & \multicolumn{2}{c|}{COGENT$(\times 10^7)$} & \multicolumn{2}{c|}{\sysname{}$(\times 10^7)$} & \multicolumn{2}{c|}{Reduction in}\\
  \cline{3-4} \cline{5-6} \cline{7-8} 
        &      & Loads & Stores & Loads & Stores & Loads & Stores\\
  \hline
  8  & 6 & 6.93 & 1.06  & 2.24 & 1.04 &  3.10 & 1.02\\
  16 & 5 & 27.8 & 6.29  & 11.9 & 2.48 &  2.33 & 2.54\\
  32 & 4 & 27.7 & 10.4  & 20.2 & 3.32 &  1.37 & 3.13\\
  64 & 3 & 6.85 & 4.71  & 3.97 & 1.48 &  1.72 & 3.18\\  
  \hline
\end{tabular}
\caption{Shared memory load and store transactions in \sysname{} and COGENT, and reduction in transactions for $\XM$ = 1024 and diverse values of $\FP$ and $\N$.
\label{fig:sh-txs}}
\end{table}

\begin{table}[t]
  \small
\begin{tabular}{|c|c|r|r|r|r|r|r|r|r|}
  \hline
  $\FP$ & $\N$ & \multicolumn{2}{c|}{\sysname{}} & \multicolumn{2}{c|}{COGENT} & \multicolumn{2}{c|}{GPyTorch} \\
  \cline{3-4} \cline{5-6} \cline{7-8}
  && Float & Double & Float & Double &Float &Double \\
  \hline
  8  & 8 & 3.90 & 1.80 & 0.67 & 0.26 & 0.26 & 0.13\\
  16 & 6 & 6.17 & 3.20 & 1.98 & 0.91 & 0.46 & 0.21\\
  32 & 5 & 7.75 & 3.88 & 5.38 & 2.26 & 1.36 & 0.64\\
  64 & 4 & 11.0 & 5.40 & 7.98 & 3.40 & 2.70 & 1.29\\
  \hline
\end{tabular}
\caption{Achieved TFLOPS of GPyTorch, COGENT, and \sysname{} for float and double with $\XM$ = 16 and largest $\FP^\N$.
\label{tab:small-m-float-double}}
\end{table}

\spara{Small $\XM$ and Double Type}
Table~\ref{tab:small-m-float-double} shows the performance for float and double types with $\XM=16$.
\sysname{} provides up to 13.4$\times$ speedup for float and up to 15.24$\times$ for double over GPyTorch. 
\sysname{} also runs up to 5.82$\times$ and 6.92$\times$ faster than COGENT for float and double respectively.

In summary, \sysname{} provides significant speedup over baselines for a diverse mix of matrix sizes and data type.

\subsubsection{Real World Dataset}
\label{sec:eval:singlegpu}
\begin{figure}[t]
  \includegraphics[scale=0.65]{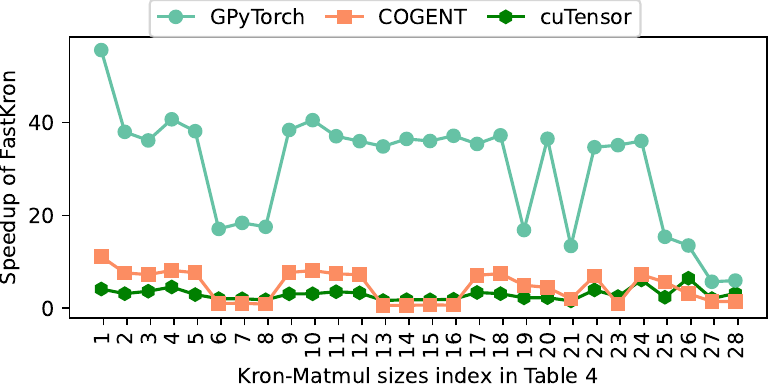}
  \caption{Speedup of \sysname{} over GPyTorch and COGENT on real-world Kron-Matmul sizes.
  The x-axis shows id of each problem size of Table~\ref{tab:results:sizes}.
  \label{fig:results:real-world-single-gpu}}
\end{figure}

We now perform experiments on the real world Kron-Matmul sizes used in machine learning compression~\cite{kru, Doping}, scientific computations~\cite{HyPA}, graph computations~\cite{Kronecker-Graphs}, computational biology~\cite{PLOS-Kron}, drugs~\cite{Viljanen2022}, and gaussian process kernels~\cite{KISS-GP-1, KISS-GP-2, skip, love, mtgpr}.
Table~\ref{tab:results:sizes} shows that our dataset contains 28 diverse cases with odd and non-power of two values of $\XM$, factors with distinct and odd sizes, and $\N$ from 2 to 11. 
Figure~\ref{fig:results:real-world-single-gpu} shows that \sysname{} performs 5.70$\times$--40.7$\times$ faster than GPyTorch, 1.43$\times$--8.14$\times$ faster over COGENT, and 1.55$\times$--6.45$\times$ faster than cuTensor on the real world dataset.



\begin{table}[t]
\footnotesize
\begin{tabular}{|c|c|r|r|r|r|}
\hline
\textbf{ID} & \textbf{Source} & \textbf{\{\XM$_\ii$\}} & \textbf{\{${\FP_{\ii}^{\N_{\ii}}\times\FQ_{\ii}}^{\N_{\ii}}$\}} \\ \hline
\multirow{4}{*}{1--5} & \multirow{4}{*}{LSTM and RNN~\cite{kru}} & 20 & $2^7\times2^7$ \\ 
               && \{20, 50\}& $2^9\times2^9$ \\ 
               && 20 & $2^{10}\times2^{10}$ \\ 
               && 1 & $2^{11}\times2^{11}$ \\ \hline
\multirow{3}{*}{6--8} & \multirow{3}{*}{ML Compression~\cite{Doping}} & 10 & $52\times50, 65\times 20$ \\ 
                     && 50 & $32\times 8$, $64\times 128$ \\ 
                     && 10 & $52\times 65$, $50 \times 20$ \\ \hline
\multirow{2}{*}{9--16} & \multirow{2}{*}{HyPA~\cite{HyPA}} & \{4, 8, 16, 20\} & $2^9\times2^9$ \\
                 && \{4, 8, 16, 20\} & $8^3\times8^3$ \\ \hline
\multirow{3}{*}{17--19} & \multirow{3}{*}{Graphs~\cite{Kronecker-Graphs}} & 1024 & $3^7\times3^7$\\ 
                                   && 1024 & $4^7\times4^7$ \\ 
                                   && 1024 & $6^7\times6^7$ \\ \hline
\multirow{2}{*}{20--21} & \multirow{2}{*}{Biology~\cite{PLOS-Kron}} & 1 & $5^3\times5^3$, $2\times2$ \\ 
                      && 1 & $5^2\times5^2$, $2\times2$, $25\times25$ \\ \hline
\multirow{3}{*}{22--24} & \multirow{3}{*}{Drug-Targets~\cite{Viljanen2022}} & 1526 & $4^6\times4^6$ \\
                      && 156  & $8^3\times8^3$ \\
                      && 2967 & $4^7\times4^7$ \\ \hline
\multirow{4}{*}{25--28} &\multirow{4}{*}{GP~\cite{KISS-GP-1, KISS-GP-2, skip,love,mtgpr}} 
   & 16 & $8^8\times8^8$ \\
   && 16 & $16^6\times16^6$ \\
   && 16 & $32^6\times32^6$ \\
   && 16 & $64^3\times64^3$ \\
\hline
\end{tabular}
\caption{\small Real world Kron-Matmul sizes.
The first column is the id for each size and the second column is the source of these sizes.
The third column $\{\XM_\ii\}$ contains one or more values of $\XM$ for the same size of factors.
The final column $\{\FP_\ii^{\N_\ii} \times \FQ_\ii^{\N_\ii}\}$ represents $\N_\ii$ consecutive factors of the shape $\FP_\ii \times \FQ_\ii$.
\label{tab:results:sizes}}
\end{table}

\subsection{Multiple GPUs Evaluation}
\label{sec:eval:multigpu}
\begin{figure}[t]
\small
\centering
\begin{subfigure}[b]{0.56\columnwidth}
\includegraphics[scale=0.71]{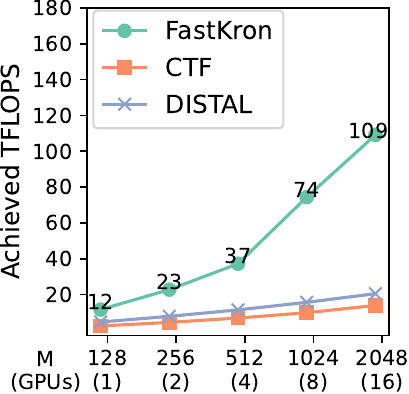}
\end{subfigure}
\hfill{}
\begin{subfigure}[b]{0.42\columnwidth}
\includegraphics[scale=0.71]{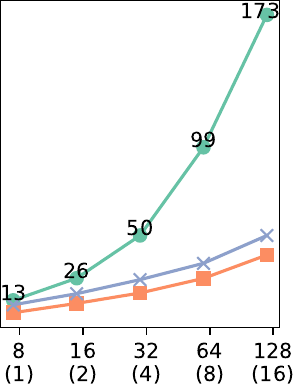}
\end{subfigure}
\caption{Weak scaling of \sysname{}, CTF, and \textsc{Distal}
on 1 to 16 GPUs with increasing $\XM$ for $\FP=64, \N=4$ (left) and $\FP=128, \N=4$ (right) with float type.
\label{fig:results:multi-gpu}}
\end{figure}

We now evaluate the multi-GPU performance of \sysname{} on 16 NVIDIA Tesla V100 GPUs.
In this experiment, we allocate all factors on all GPUs and compare \sysname{} against the following state-of-the-art distributed tensor algebra frameworks:

\spara{Cyclops Tensor Framework} (CTF)~\cite{ctf} implements distributed tensor matrix multiply as a series of distributed Matmuls and transposes.
Thus, our implementation of the distributed version of the shuffle algorithm in CTF uses a series of distributed tensor matrix multiplies.

\spara{\textsc{Distal}}~\cite{distal} allows a user to manually specify a distributed schedule for the given tensor algebra computation.
We implemented each iteration of the FTMMT algorithm in \textsc{Distal}.
Our schedule of the algorithm follows the same distribution as \sysname{}, i.e., divide $\XM$ and $\XK$ by $\sqrt{G}$.
However, it is not possible to specify our distributed Kron-Matmul in \textsc{Distal} because \textsc{Distal} communicates intermediate after every sliced multiplication.

\subsubsection{Results}
Figure~\ref{fig:results:multi-gpu} shows the weak scaling (memory per GPU remains constant) performance of all systems with increasing $\XM$ for $\FP = 64, \N = 4$ and $\FP=128, \N=4$.
We chose these values of $\FP$ and $\N$ because they provide maximum FLOPs per GPU.
\sysname{} provides speedup of 7.85$\times$ over CTF and 5.33$\times$ over \textsc{Distal} at 16 GPUs.
Moreover, for 16 GPUs, \sysname{} reaches 69\% of the maximum FLOPs.
\textsc{Distal} performs better than CTF because CTF performs distributed transposes, which \textsc{Distal} avoids.
\sysname{} performs better than both \textsc{Distal} and CTF because \sysname{} minimizes the communication volume by performing multiple sliced multiplications on each GPU before communicating their intermediates to obtain the full intermediate.
Thus, \sysname{} is an efficient distributed Kron-Matmul engine.

\subsection{Case Study: Fast Training of Gaussian Processes}
\label{sec:end-to-end-eval}
\emph{Gaussian Processes} (GPs) are a class of machine learning 
models that provide predictions with uncertainty and interpretability~\cite{10.7551/mitpress/3206.001.0001}.
GPs contain a kernel matrix $\textbf{K}$ and represent the training dataset of \XM{} points as a vector $\V$ of length \XM{}.
The training process of GPs computes $(\textbf{K})^{-1}\V$ ~\cite{gpbook, KISS-GP-1}, which can be expensive for large values of $\XM{}$.
Structured Kernel Interpolation (SKI)~\cite{KISS-GP-1, KISS-GP-2} is a GP that interpolates the kernel matrix as 
$\textbf{W} (\textbf{K}^1 \kron \textbf{K}^2 \ldots \textbf{K}^\text{N})\textbf{W}^\text{T}$, 
where $\textbf{W}_{\XM{} \times \FP^\N}$ is an interpolation weight matrix 
and 
$\textbf{K}^\ii_{\FP \times \FP}$ is a Kronecker kernel matrix.
Using the conjugate gradient algorithm, the inverse computation is done using a series of Kron-Matmuls of $\V$ and $\kron_\ii \textbf{K}^\ii$.
Thus, Kron-Matmul is a key process in training SKI and its variants, SKIP~\cite{skip} and LOVE~\cite{love}.
Also, large values of $\FP$ and $\N$ improve the accuracy of GPs.

We integrated \sysname{} in GPyTorch to accelerate Kron-Matmul and use it to evaluate the reduction in training time of SKI, SKIP, and LOVE on the UCI dataset~\cite{ucidataset} with 150 to 3$\times 10^5$ points.
We set the conjugate gradient method to consider 16 samples, i.e., \XM{} = 16, and runs for 10 iterations in each epoch.
These datasets and parameters have been used in prior works~\cite{KISS-GP-1,KISS-GP-2, skip, love}.
We perform experiments on the highest value of $\FP$ that can be allocated in the GPU memory for each dataset.
Table~\ref{tab:gp-integration} shows that integrating \sysname{} in GPyTorch provides a speedup of up to 1.95$\times$ on a single GPU and up to 6.20$\times$ on 16 GPUs when training GPs.
GPyTorch does not support multi-GPU execution for these GPs and thus executes several other operations on a single GPU, leading to a speedup increase of up to 3.33$\times$ with 16 GPUs over 1 GPU.
Thus, integrating \sysname{} enables faster training of GPs on larger kernel matrices.

\begin{table}[t]
  \small
  \begin{tabular}{|@{}c@{}|c|c|r|r|r|r|r|r|}
    \hline
    Dataset & $\FP^\N$ & \multicolumn{3}{c|}{Speedup on 1 GPU} & \multicolumn{3}{c|}{Speedup on 16 GPUs}\\
    \cline{3-5} \cline{6-8}
    &       &      SKI    & SKIP         & LOVE              & SKI    & SKIP    & LOVE\\
    \hline
    autompg& $8^7$  &1.1$\times$ &1.1$\times$  & 1.2$\times$      & 1.3$\times$ & 1.3$\times$ & 1.5$\times$\\
    kin40k& $8^8$  &1.5$\times$ &1.3$\times$  & 1.2$\times$      & 3.1$\times$ & 1.8$\times$ & 1.6$\times$\\
    \hline
    airfoil& $16^5$ &1.1$\times$ &1.1$\times$  & 1.3$\times$      & 1.2$\times$ & 1.2$\times$ & 1.5$\times$\\
    yacht& $16^6$ &1.8$\times$ &1.7$\times$  & 1.9$\times$      & 3.8$\times$ & 3.3$\times$ & 5.2$\times$\\
    \hline
    servo& $32^4$ &1.1$\times$ &1.1$\times$  & 1.2$\times$      & 1.3$\times$ & 1.2$\times$ & 1.5$\times$\\
    airfoil& $32^5$ &1.8$\times$ &1.8$\times$  & 1.8$\times$      & 6.2$\times$ & 4.9$\times$ & 5.0$\times$\\
    \hline
    3droad & $64^3$ &1.1$\times$ &1.1$\times$  & 1.2$\times$      & 1.2$\times$ & 1.2$\times$ & 1.1$\times$\\
    servo & $64^4$ &2.1$\times$ &2.0$\times$  & 2.2$\times$      & 4.5$\times$ & 3.8$\times$ & 5.4$\times$\\
    \hline
  \end{tabular}
  \caption{Speedups in training GPs on real world datasets after integrating \sysname{} in GPyTorch over vanilla GPyTorch.\label{tab:gp-integration}}
  \vspace{-2em}
\end{table}
\balance

\section{Related Work}

\spara{Tensor Contractions}
The traditional way to execute tensor contractions first transpose input tensors to a valid Matmul, and then transpose the output to required tensor.
Since these transposes are expensive, multiple works have developed efficient transpose routines for CPUs~\cite{hptt, ttc}, and GPUs~\cite{talsh, ttc, cutt, ttlg}.
TTC~\cite{ttc} and TTLG~\cite{ttlg} are compilers for transpose routines.
TAL\_SH~\cite{talsh} uses the state-of-the-art cuTT~\cite{cutt} library for efficient transpose on GPUs.
Several works avoid the transpose and directly perform tensor contraction~\cite{tblis, GETT, tcgen, nwchem, 7839684}.
TBLIS~\cite{tblis} fuses transpose with BLIS~\cite{blis} Matmul kernels on CPUs.
GETT~\cite{GETT} uses a highly tuned macro kernel where its operands reside in the cache hierarchy.
CUTLASS~\cite{cutlass} and cuTensor~\cite{cutensor} extends GETT approach to GPUs.
Nelson et. al~\cite{7349652} and Patabandi et. al.~\cite{10.1145/3460945.3464955} uses machine learning to tune tile size parameters of a tensor algebra GPU kernel.
COGENT~\cite{tcgen} improves over these approaches by generating a specialized kernel and tile sizes for tensor contractions for GPUs.
Thus, these works can efficiently execute each multiplication with a Kronecker factor in the FTMMT algorithm.
However, unlike \sysname{}, these works do not optimize for memory accesses across multiplications in the algorithm.
Kim et. al.\cite{tcgen-fusion} improves tensor contractions for coupled cluster methods in quantum chemistry by fusing multiple contractions.
However, their approach performs transpose in shared memory and these tensor contractions are different from contractions in Kron-Matmul.


\spara{Kronecker Matrix-Matrix Multiplication} 
GPytorch~\cite{gpytorch} and PyKronecker~\cite{pykronecker} are two state-of-the-art single-GPU implementations for the shuffle algorithm~\cite{shuffle-algo}.
Dayar and Orhan~\cite{doi:10.1137/140980326} presents an improvement to the shuffle algorithm for Kronecker matrix-vector products.
Fackler~\cite{10.1016/j.cam.2003.10.010} proposed an algorithm similar to FTMMT that avoids the transpose by representing the input matrix as a tensor.
We use COGENT~\cite{tcgen,tcgen-fusion} as a baseline for this algorithm.
\sysname{} improves over these implementations by avoiding transpose and using optimizations, such as, shift caching and fusion of iterations.
Moreover, \sysname{} provides a distributed algorithm while above systems are only for single node.

\spara{Optimizing Small and Skinny Matmul}
Many works has optimized Matrix Multiplication and Matrix-Vector Multiplication computations on small and skinny matrices on GPUs~\cite{cpe.4705, kblas,tsm2x}.
He et al.~\cite{cpe.4705} proposes an optimal warp allocation strategy for matrix-vector multiplication.  
KBLAS~\cite{kblas} uses double-buffering to overlap data motion with computation to optimize matrix-vector multiplication. 
TSM2X~\cite{tsm2x} optimizes GEMM of rectangular matrices with small matrices.
These techniques can improve the Matmul part in the shuffle algorithm but will still suffer from high transpose cost.
However, \sysname{} avoids transpose operations and also provides an efficient multi-GPU execution.

\spara{Distributed Tensor Algebra}
Cannon~\cite{cannon,cannon-ics17} and SUMMA \cite{summa} are one of the first algorithms for distributed Matmul.
Solomonik et. al.~\cite{edgar25D} presented a 2.5D algorithm that distributes the summation dimension and CARMA~\cite{recursive} is a recursive algorithm.
COSMA~\cite{cosma} is an near I/O-optimal algorithm for distributed Matmul.
Rajbhandari et. al.~\cite{7013018} presents a communication-optimal algorithm for distributed tensor contraction.
Cyclops Tensor Framework (CTF)~\cite{ctf} and DISTAL~\cite{distal} are two state-of-the-art distributed tensor algebra systems.
CTF executes tensor contractions as a series of distributed transposes and Matmuls, while DISTAL allows fusion of the transpose with contraction to perform better than CTF.
For Kron-Matmul, both approaches communicates intermediate for each iteration,
while \sysname{} minimizes the communication by performing multiple sliced multiplications on each GPU and communicate intermediate of last multiplication.

\spara{Distributed Computations} 
Distributed Halide~\cite{distributed-halide} extends Halide with scheduling primitives for distributing dimensions of loops.
Recent works~\cite{dcuda, coconet} supports overlapping CUDA computations with communication.
However, Kron-Matmul algorithm cannot be represented in these frameworks.


\section{Conclusion}
In this paper, we proposed a novel algorithm for Kron-Matmul, which is not based on existing linear algebra operations.
This advantage enabled us to develop new optimizations for Kron-Matmul implementations on GPUs.
Experimental results demonstrates that our implementation outperforms state-of-the-art techniques on both single and multiple GPUs.

\clearpage

\bibliographystyle{ACM-Reference-Format}
\bibliography{paper}

\newpage

\appendix
\section{Artifact Appendix}
The artifact~\cite{artifact} contains implementation of \sysname{} and scripts to reproduce our key results.
The artifact provides a Dockerfile, which contains all prerequisites installed.
Latest source code is available at \url{https://github.com/abhijangda/fastkron}.

\subsection{Hardware}
\sysname{} supports both systems with a single NVIDIA GPU and multiple NVIDIA GPUs.
In our experiments we use a DGX-2 machine with 16 NVIDIA Tesla V100 GPUs connected using NVLINK 2.

\subsection{Docker Container}
Download the artifact zip file from \cite{artifact}, unzip it, and create the container.
\begin{lstlisting}[basicstyle=\small\ttfamily]
unzip fastkron-ppopp-24-ae.zip
cd fastkron-ae
docker build -t fastkron-ppopp-24-ae .
docker run -it --gpus all fastkron-ppopp-24-ae
\end{lstlisting}
Check if PyTorch supports CUDA:
\begin{lstlisting}[basicstyle=\small\ttfamily]
python
>>> import torch
>>> torch.cuda.is_available()
True
\end{lstlisting}

\subsection{Getting Started}
We will now build FastKron and execute tests. 
In the container, the FastKron directory is available at \texttt{/fastkron} and the benchmark infrastructure is in \texttt{/fastkron-benchmarks}.
\spara{Setup CMake} Setup CMake inside FastKron Directory

\begin{lstlisting}[basicstyle=\small\ttfamily]
mkdir /fastKron/build
cd /fastKron/build
cmake ..
\end{lstlisting}

\spara{Single GPU Test} 
We can execute one of the single GPU tests as below:

\begin{lstlisting}[basicstyle=\small\ttfamily]
make gen-single-gpu-kernels 
make run-single-gpu-no-fusion-tests -j
\end{lstlisting}

\spara{Multi GPU Test}
We can execute one of the multi GPU tests as below:

\begin{lstlisting}[basicstyle=\small\ttfamily]
make gen-multi-gpu-tests-kernel 
make run-multi-gpu-nccl-no-fusion-tests -j
\end{lstlisting}

\spara{Execute all Tests (Optional)} 
We can execute all tests from the FastKron directory
\begin{lstlisting}[basicstyle=\small\ttfamily]
cd /fastkron
python tests/run-tests.py 
\end{lstlisting}
If all above tests run fine and do not give any error then we have successfully setup the benchmarking.

\subsection{Step by Step Instructions}
We will now reproduce results in Figure~\ref{fig:results:single-gpu-flops}, Table~\ref{tab:small-m-float-double}, Figure~\ref{fig:results:real-world-single-gpu}, Figure~\ref{fig:results:multi-gpu}, and Table~\ref{tab:gp-integration}. These commands generate figures as PDF in the benchmarks directory and table as CSV in the benchmarks directory. 
\\Change to the benchmark directory:

\begin{lstlisting}[basicstyle=\small\ttfamily]
cd /fastkron-benchmarks
\end{lstlisting}

\spara{Figure 9} [Time 30 mins] Generate \texttt{Figure-9.pdf} in the benchmarks directory by executing:

\begin{lstlisting}[basicstyle=\small\ttfamily]
python run_benchmarks.py -fk-dir /fastkron\
       -fk-bench-dir /fastkron-benchmarks \
       -bench Figure-9
make Figure-9.pdf
\end{lstlisting}

\spara{Table 3} [Time 15 mins] Generate \texttt{Table-3-float.csv} for Float type and \texttt{Table3-double.csv} for Double type in the benchmarks directory by executing: 

\begin{lstlisting}[basicstyle=\small\ttfamily]
python run_benchmarks.py -fk-dir /fastkron\
       -fk-bench-dir /fastkron-benchmarks \
       -bench Table-3
\end{lstlisting}

\spara{Figure 10} [Time 40 mins] Generate \texttt{Figure-10.pdf} in the benchmarks directory by executing:

\begin{lstlisting}[basicstyle=\small\ttfamily]
python run_benchmarks.py -fk-dir /fastkron\
       -fk-bench-dir /fastkron-benchmarks \
       -bench Figure-10
make Figure-10.pdf
\end{lstlisting}

\spara{Figure 11} [Time 40 mins] Generate \texttt{Figure-11-64.pdf} and \texttt{Figure-11-128.pdf}:

\begin{lstlisting}[basicstyle=\small\ttfamily]
python run_benchmarks.py -fk-dir /fastkron\
       -fk-bench-dir /fastkron-benchmarks \
       -bench Figure-11
make Figure-11-64.pdf Figure-11-128.pdf
\end{lstlisting}

\spara{Table 5} [Time 30 mins] Generate \texttt{Table-5.csv} by executing:
\begin{lstlisting}[basicstyle=\small\ttfamily]
python gps-Table-5.py ./uci 10
\end{lstlisting}

\end{document}